\documentclass{sig-alternate-05-2015}
\pdfoutput=1

\setcopyright{acmlicensed}

\doi{10.475/123_4}

\isbn{123-4567-24-567/08/06}

\conferenceinfo{CIKM '16}{October 24--28, 2016, Indianapolis, USA}

\acmPrice{\$15.00}

\usepackage{times}
\usepackage{latexsym}
\usepackage{tabularx}
\usepackage{booktabs}
\usepackage{amsmath}
\usepackage{amssymb}
\usepackage{multirow}
\usepackage{url}
\usepackage[square,comma,numbers,sort&compress,sectionbib]{natbib}
\usepackage[usenames,dvipsnames]{xcolor}
\usepackage[normalem]{ulem}
\usepackage{mathcomp}
\usepackage{setspace}
\usepackage{paralist}
\usepackage{enumitem}
\setlist{nosep}
\usepackage[group-separator={,},group-minimum-digits=4]{siunitx}

\usepackage{subfigure}

\usepackage{graphicx}
\usepackage{algorithm}
\usepackage{algorithmic}

\newcommand{\OurApproach}{EIDF}

\renewcommand{\paragraph}[1]{\smallskip\noindent\textbf{#1.}}

\newcommand{\avg}{\mathop{\mathrm{avg}}}

\begin{document}

\title{Document Filtering for Long-tail Entities}

\numberofauthors{1}

\author{
	\alignauthor
	\hfill
	\begin{tabular}{c}
		Ridho Reinanda\raisebox{4pt}{$\dag$} \\
		\email{r.reinanda@uva.nl}
	\end{tabular}
	\hfill
	\begin{tabular}{c}
		Edgar Meij\raisebox{4pt}{$\ddagger$} \\
			\email{edgar.meij@acm.org}
	\end{tabular}
	\hfill
	\begin{tabular}{c}
		Maarten de Rijke\raisebox{4pt}{$\dag$} \\
		\email{derijke@uva.nl}	
	\end{tabular}		
    \hfill\mbox{}\\[1.1ex]
    \affaddr{\raisebox{4pt}{$\dag$}~University of Amsterdam, Amsterdam, The Netherlands}\\
    \affaddr{\raisebox{4pt}{$\ddagger$}~Bloomberg L.P., London, United Kingdom}
}

\CopyrightYear{2016}
\setcopyright{acmlicensed}
\conferenceinfo{CIKM'16 ,}{October 24 - 28, 2016, Indianapolis, IN, USA}
\isbn{978-1-4503-4073-1/16/10}\acmPrice{\$15.00}
\doi{http://dx.doi.org/10.1145/2983323.2983728}

\maketitle
 
\begin{abstract}
Filtering relevant documents with respect to entities is an essential task in the context of knowledge base construction and maintenance. It entails processing a time-ordered stream of documents that might be relevant to an entity in order to select only those that contain vital information. State-of-the-art approaches to document filtering for popular entities are entity-dependent: they rely on and are also trained on the specifics of differentiating features for each specific entity. Moreover, these approaches tend to use so-called \emph{extrinsic} information such as Wikipedia page views and related entities which is typically only available only for popular head entities.
Entity-dependent approaches based on such signals are therefore ill-suited as filtering methods for long-tail entities.
In this paper we propose a document filtering method for long-tail entities that is \emph{entity-independent} and thus also generalizes to unseen or rarely seen entities. It is based on intrinsic features, i.e., features that are derived from the documents in which the entities are mentioned. We propose a set of features that capture informativeness, en\-tity-saliency, and timeliness. In particular, we introduce features based on entity aspect similarities, relation patterns, and temporal expressions and combine these with standard features for document filtering. 
Experiments following the TREC KBA 2014 setup on a publicly available dataset show that our model is able to improve the filtering performance for long-tail entities over several baselines. Results of applying the model to unseen entities are promising, indicating that the model is able to learn the general characteristics of a vital document. The overall performance across all entities---i.e., not just long-tail entities---improves upon the state-of-the-art without depending on any entity-specific training data.
\end{abstract}

%
%


%
%

%
%

\vspace*{-.25\baselineskip}


\section{Introduction}
\label{section:introduction}

A knowledge base contains information about entities, their attributes, and their relationships. Modern search engines rely on knowledge bases for query understanding, question answering, and document enrichment \cite{Balasubramanian2010, Voskarides2015, Pantel2011}.
Knowledge-base construction, either based on web data or on a domain-specific collection of documents, is the cornerstone that supports a large number of downstream tasks. In this paper, we consider the task of entity-centric document filtering, which was first introduced at the TREC KBA evaluation campaign~\cite{Frank2012}. Given an entity, the task is to identify documents that are relevant and vital for enhancing a knowledge base entry of the entity given a stream of incoming documents.

To address this task, a series of \emph{entity-dependent} and \emph{entity-in\-dependent} approaches have been developed over the years. Entity-dependent approaches use features that rely on the speci\-fics of the entity on which they are trained and thus do not generalize to unseen entities. Such methods include approaches that learn a set of keywords related to each entity and utilize these keywords for query expansion and document scoring~\cite{Liu2013, Dietz2013} as well as text-classification-based approaches that build a classifier with bag-of-word features for each entity~\cite{Frank2012}. Signals such as Wikipedia page views and query trends have been shown to be effective, since they usually hint at changes happening around an entity~\cite{Balog2013}; these signals are typically available for popular entities but when working with long-tail entities, challenges akin to the cold-start problem arise. In other words, features extracted from and working for popular entities may simply not be available for long-tail entities.

In this paper, we are particularly interested in filtering documents for long-tail entities. Such entities have limited or even no external knowledge base profile to begin with. Other extrinsic resources may be sparse or absent too. This makes an entity-dependent document filtering approach a poor fit for long-tail entities. Rather than learning the specifics of each entity, \emph{entity-indepen\-dent} approaches to document filtering aim to learn the characteristics of documents suitable for updating a knowledge base profile by utilizing signals from the documents, the initial profile of the entity (if present), and relationships between entities and documents~\cite{Balog2013, Wang2013, Wang2015}. While entity-dependent approaches might be able to capture the distributions of features for each entity better, entity-indepen\-dent approaches have the distinct advantage of being applicable to unseen entities, i.e., entities not found in the training data. As an aside, entity-independent methods avoid the cost of building a model for each entity which is simply not practical for an actual production-scale knowledge base acceleration system.

Our main hypothesis is that a rich set of \emph{intrinsic} features, based on aspects, relations, and the timeliness of the facts or events mentioned in the documents that are relevant for a given long-tail entity, is beneficial for document filtering for such entities.
We consider a rich set of features based on the notion of \emph{informativeness}, \emph{entity-saliency}, and \emph{timeliness}. The intuition is that a document~\begin{inparaenum}[(1)]\item that contains a rich set of facts in a timely manner, and \item in which the entity is prominent\end{inparaenum}{} makes a good candidate for enriching a knowledge base profile. 
To capture informativeness, we rely on three sources: generic Wikipedia section headings, open relations, and schematized relations in the document. To capture entity-saliency, we consider the prominence of an entity with respect to other entities mentioned in the document. 
To capture timeliness, we consider the time expressions mentioned in a document. We use these features with other basic features to train an entity-independent model for document filtering for long-tail entities.

Our main contributions can be summarized as follows:
\begin{inparaenum}[(1)]
	\item We propose a competitive entity-independent model for document filtering for long-tail entities with rich feature sets designed to capture informativeness, entity-saliency, and timeliness.
	\item We provide an in-depth analysis of document filtering for knowledge base acceleration for long-tail entities.
\end{inparaenum}

\section{Related Work}
\label{section:related-work}

We review related work on document filtering in the TREC KBA track and other settings as well as related work on entity profiling.

\subsection{Document filtering}

The main approaches to KBA (and TREC KBA in particular) can be divided into entity-dependent and entity-independent approaches.
%
When TREC KBA was first organized in 2012, many methods relied on \emph{entity-dependent}, highly-supervised approaches utilizing related entities and bag of word features~\cite{Frank2012}. Here, the training data is typically used to identify keywords and/or related entities, in order to classify the documents in the test data. Later on, entity-independent models which rely less on the specifics of each entity emerge.  
\citet{Balog2013} propose one such \emph{entity-independent} approach. They study two multi-step classification methods for the stream filtering task. Their models start with an entity identification component based on alternate names from Wikipedia. They introduce a set of features that have commonly been used in subsequent TREC KBA campaigns. \citet{Balog2012} also compare classification and ranking approaches for this task; ranking outperforms classification on all evaluation settings and metrics on the TREC KBA 2012 dataset. Their analysis reveals that a ranking-based approach has more potential for future improvements. 

Along this line of work, \citet{Bonnefoy2013} introduce a weakly-supervised, entity-independent detection of the central documents in a stream. \citet{Zhou2013} study the problem of learning entity-centric document filtering based on a small number of training entities. They are particularly interested in the challenge of transferring keyword importance from training entities to entities in the test set. They propose novel meta-features to map keywords from different entities and contrast two different models: linear mapping and boost mapping. 

\citet{Wang2013} adopt the features introduced in \cite{Balog2013} and introduce additional citation-based features, experimenting with different classification and ranking-based models.  They achieve the best performance for vital documents filtering in KBA 2013 with a classification-based approach. \citet{Liu2013} present a related entity-based approach. They pool related entities from the profile page of target entity and estimate the weight of each related entity with respect to the query entity. They then apply the weighted related entities to estimate confidence scores of streaming documents and explore various ways of weighting the related entities. \citet{Dietz2013} also propose a query expansion-based approach on relevant entities from the KB. They do not address the novelty aspects of the task, however, and evaluate a memory-less method where predictions are not influenced by predictions on previous time intervals.

\citet{Wang2015} propose a novel discriminative mixture model based by introducing a latent entity class layer to model the correlations between entities and latent entity classes. 
They achieve increased performance by inferring latent classes of entities and learning the appropriate feature weights for each latent class, as shown by experiments on the TREC KBA 2013 dataset. Later on, \citet{Wang2015b} introduce a latent document filtering model for cumulative citation recommendation in which they infer different latent classes and learn the appropriate feature weights for each latent class.

\citet{Gebremeskel2014} perform an in-depth analysis of the main factors that affect the recall of document filtering on the TREC KBA 2013 corpus. They investigate the impact of choices for corpus cleansing, entity profile construction, entity type, document type, and relevance grading. They identify and characterize citation-worthy documents that do not pass the filtering stage by examining their contents and find that this can be caused by cleansing issues, incomplete name variants, or unclear assessments reasons.

In contrast with previous years, TREC KBA 2014 focused on long-tail entities and less than half of the entities in the test set that year have a Wikipedia profile~\cite{Frank2014}. \citet{Jiang2014} achieved the best performance using an entity-dependent approach which uses time range, temporal, profession, and action pattern features. Another notable approach within that year summarizes all information known about an entity so far in a low-dimensional embedding \cite{Cano2014}.

Next, we describe two tasks that are different but closely related to the document filtering setting of TREC KBA. Document filtering has been a traditional task in TREC, in the form of Topic Detection and Tracking (TDT)~\citep{Allan2002}. TDT constitutes a body of research and evaluation paradigm that address event-based organization of broadcast news. The goal of TDT is to break the text down into individual news stories, to monitor the stories for events that have not been seen before, and to gather stories into groups that each discuss a single news topic.

\citet{Dunietz2014} introduce the entity salience task, that is given a document $d$, decide whether entity $e \in E_d$ is salient, i.e., a major talking point of the document. This task is similar to document filtering without the requirement of having the documents mentioning timely facts.

\subsection{Entity profiling}

Next, we turn to discovering entity-oriented pieces of information within a text.
\citet{Fetahu2015} propose a two-stage supervised approach for suggesting news articles to entity pages for a given state of Wikipedia. First, they suggest news articles to Wikipedia entities (article-entity placement), relying on a rich set of features which take into account the salience and relative authority of entities, and the novelty of news articles to entity pages. Next, they determine the exact section in the entity page for the input article (predicting the correct section for the article) guided by what they call class-based section templates. \citet{Banerjee2015} explore the task of automatically expanding Wikipedia stubs. They introduce a model that assigns web content to a Wikipedia section and then perform abstractive summarization to generate section-specific summaries for the Wikipedia stubs.

\citet{Taneva2013} propose an approach that automatically compiles salient information about entities in order to ease knowledge bases maintenance. They compile highly-informative, concise ``gems'' about entities, identifying salient pieces of text of variable granularity using a budget-constrained optimization problem, which decides which sub-pieces of an input text should be selected for the final result. 
\citet{Li2011} propose a novel approach to automatically generate aspect-oriented summaries from multiple documents. They first introduce an event-aspect LDA model to cluster sentences into aspects and then use LexRank to rank the sentences in each cluster, employing Integer Linear Programming for sentence selection. 
\citet{Song2012} present a model to summarize a query's results using distinct aspects. For this they introduce the notion of ``composite queries'' that are used for providing additional information for a query and its aspects, comparatively mining the search results of different component queries. \citet{Balasubramanian2009} propose a method to generate entity-specific topic pages as an alternative to regular search results. \citet{Cheng2015} study the task of generating compact structured summaries. 
%
\citet{reinanda-mining-2015} mine entity aspects, common information needs around entities, from query logs, while  \citet{reinanda-prior-informed-2014} focus on establishing temporal extents of entity relations, which can be useful for updating sections of entity profiles that are temporal in nature.

\smallskip\noindent%
Our work is different in the following ways. First, we focus on the vital document filtering for long tail entities specifically. Next, we introduce a rich set features for identifying vital documents based on the notion \emph{informativeness}, \emph{entity-saliency}, and \emph{timeliness} of the documents. Last, we apply these rich features to train an entity-independent model for vital document filtering.

\begin{table}[t]
\centering
 \caption{
   Glossary of the main notation used in this paper.  }
 \label{tbl:gloss}
\begin{tabularx}{\linewidth}{lX}
  \toprule
  \textbf{Symbol} & \textbf{Gloss}\\
  \midrule
   $S$ & Stream of documents\\
   $d$ & a document\\
   $e$ & an entity\\
   $p$ & a profile of an entity\\
   $a$ & an aspect of an entity\\
  \bottomrule
  \end{tabularx}
\end{table}%

\section{Problem Definition}
\label{section:problem}

In this paper, we study the problem of identifying documents that contain vital information to add to a knowledge base. We formalize the task as follows.
Given an entity $e$ and a stream of documents $S$, we have to decide for each document $d_e \in S$ that mentions $e$ whether it is vital for improving a knowledge base profile $p_e$ of entity $e$. More formally, we have to estimate:
\begin{equation}
	P(rel\mid d_e,e),
	\label{eq:problem}
\end{equation} 
where $rel$ is the 
relevance
of document $d_e$ with respect to entity $e$. A document is considered \emph{vital} if it can enhance the current knowledge base profile of that entity, for instance by mentioning a fact about the entity within a short window of time of the actual emergence of the new fact.
%
Note that a profile $p_e$ is a textual description of an entity (i.e., not a structured object), such as a Wikipedia page or any other web page providing a description of the entity at a certain point in time.



\section{Method}
\label{section:method}

In this section, we describe our general approach to perform document filtering. We consider several intrinsic properties of a document that will help to detect vital documents.
In particular, we consider the following dimensions:

\begin{itemize}
	\item{\textbf{Informativeness}} -- a document $d$ that is rich in facts is likely to be vital.
	\item{\textbf{Entity-saliency}} -- a document $d$ in which an entity $e$ is salient among the set of entities $E$ occurring in $d$ is likely to be vital.
	\item{\textbf{Timeliness}} -- a document $d$ that contains and discusses a timely event (with respect to document creation time or classification time) is likely to be vital. 
\end{itemize}


\noindent
We hypothesize that not all of these properties need to be satisfied for a document to be considered vital, i.e., some combination of features derived from these properties and other basic features for document filtering would apply in different cases. 


\subsection{Intrinsic features}

Below, we describe the intrinsic features derived to capture the three dimensions described above and how these features are used to operationalize Eq.~\ref{eq:problem}. The features are meant to be used in combination with others that are commonly used in document filtering and that will be described below.
In the following paragraphs we describe these features; a high-level summary can be found in Table~\ref{table:features}. 

\subsubsection{Informativeness features}

Informativeness features aim to capture the richness of facts contained in a document. The intuition is that a document that contains a lot of facts, for instance in the form of relations, such as \emph{work-for}, \emph{spouse-of}, \emph{born-in}, would be more likely to be vital. We operationalize informativeness in three ways, using entity page sections in a knowledge base (e.g., Wikipedia), open relations, and schematized relations as detailed below. We denote the informativeness features as $F_I$.

\paragraph{Wikipedia aspects} We define aspects as key pieces of information with respect to an entity. The central idea here is that a vital document contains similar language as some specific sections in Wikipedia pages; cf.~\citep{fissaha-adafre-discovering-2005}. We therefore aggregate text belonging to the same Wikipedia section heading from multiple Wikipedia pages
in order to build a classifier. 
To be able to extract aspect features for a document, we first construct a bag-of-words model of aspects $A_c$ of an entity type $c$ from Wikipedia as detailed in Algorithm~\ref{algorithm1}. Here we first retrieve Wikipedia articles of \emph{all} entities belonging to the Wikipedia category $c$; our entities are filtered to be either in the \emph{Person} or \emph{Location} category. Next, we identify the section headings within the articles.  We take the $m$ most frequent section headings and, for each section heading, we remove stopwords and aggregate the contents belonging to the same heading by merging all terms that occur in the heading as an aggregated bag-of-words. We then represent each aggregated content section as a bag-of-words representation of aspect $a_k \in A$ and 
compute the cosine similarity between the candidate document $d$ and aspect $a_k$ to construct an aspect-based feature vector 
\begin{equation}
	A_k(d) = \cos(d, a_k),
\end{equation}
We refer to the vector $A_k$ as the $\mathit{ASPECTSIM}$ features in Table~\ref{table:features}.

\begin{algorithm}[t]
\caption{Building a Wikipedia aspect model.}
\label{algorithm1}
\begin{algorithmic}[1]
\REQUIRE~ Wikipedia entity category: $c$, Wikipedia articles: $W$
\ENSURE~Aspect model: $A_c$; \\

\STATE $C \gets retrieveArticles(W,c)$
\STATE $H_C \gets extractSectionHeadings(C)$
\STATE $aggregateSectionHeadings(H_C)$
\FOR { each $h \in H_C$ }
	\STATE $S_C \gets retrieveSections(H_C, h)$
	\STATE $a_s \gets combineSections(S_C)$
\ENDFOR

\end{algorithmic}
\end{algorithm}

\paragraph{Open relation extraction} Here, we use the relation phrases available from an open information extraction system, i.e., Reverb~\cite{Etzioni2011}. As an open relation extraction system, Reverb does not extract a predefined set of entity relations from text, but detects any relation-like phrases. Given a text as input, it outputs unnormalized relational patterns in the form of triples of an entity, a verb/noun phrase, and another entity. As another feature, we utilize the relational patterns generated by Reverb from the ClueWeb09 corpus~\cite{Fader2011}.
Algorithm~\ref{algorithm2}  details our procedure to generate a list of open relation phrases from this output.
Due to the large number of patterns and limited amount of training data, it is not feasible to use all of these patterns as features. Therefore, we select popular phrases out of all available patterns. To this end, we first cluster the relation phrases based on their lemmatized form, obtaining grouped patterns $G$. Then, we estimate the importance of each pattern group $g \in G$ based on their aggregated count in the ClueWeb09 corpus. That is, we sum the occurrence $c_p$ of each pattern $p$ as the count of group $g$, obtaining $c_g$. Finally, we select the $n$ most frequent relation phrases.
We compute the feature vector by splitting a document into sentences and, for each relation phrase $R$ compiled in the previous step, we generate a feature vector containing the counts:
\begin{equation}
	R_k(d) = count(d, r_k),
\end{equation}
where $count(d,r)$ returns the count of any instances of open relation pattern $r$ in the document $d$. We refer to the vector $R_k$ as the $RELOPEN$ features in Table~\ref{table:features}.

\begin{algorithm}[t]
\caption{Selecting open relation phrase patterns.}
\label{algorithm2}
\begin{algorithmic}[1]
\REQUIRE~ Open relation phrases: $P$, Corpus $C$
\ENSURE~Ranked open relations model: $R$; \\

\STATE $G \gets groupPhrasesByLemma(P)$
\FOR { each $g \in G$ }
	\FOR { each $p \in g$ }
		\STATE $c_p \gets getCount(C,p)$
	\ENDFOR
	\STATE $c_g \gets c_g + c_p$
\ENDFOR

\STATE $R \gets selectTopk(G, c)$
\end{algorithmic}
\end{algorithm}

\begin{table*}[t]
	\caption{Features for document filtering, for an entity $e$ and/or document $d$. The last column indicates the value types of the features: $N$ for numerical features and $C$ for categorical features.}
	\label{table:features}
	\centering
	\begin{tabularx}{\linewidth}{lXllc}
		\toprule
		\bf Feature & \bf Description & \bf Source & \bf Type & \bf Value \\
		\midrule
		$\mathit{SRC}(d)$ & Document source/type & \cite{Balog2013} & basic & N \\
		$\mathit{LANG}(d)$ & Document language & \cite{Balog2013} & basic & N \\
		$\mathit{REL}(e)$ & Number of of related entities of $e$ & \cite{Balog2013} & basic & N \\
		$\mathit{DOCREL}(e)$ & Number of of related entities of $e$ in $d$ & \cite{Balog2013} & basic & N \\
		$\mathit{NUMFULL}(d,e)$ & Number of mentions of $e$ in $d$ & \cite{Balog2013} & basic & N \\
		$\mathit{DOCREL}(d,e)$ & Number of of related entities of $e$ in $d$ & \cite{Balog2013} & basic & N \\
		$\mathit{NUMPARTIAL}(d,e)$ & Number of partial mentions of $e$ in $d$ & \cite{Balog2013} & basic & N \\
		$\mathit{FPOSFULL}(d,e)$ & First position of full mention of $e$ in $d$  & \cite{Balog2013} & basic & N \\
		$\mathit{LPOSPART}(d,e)$ & Last position of partial mention of $e$ in $d$  & \cite{Balog2013} & basic & N \\
		$\mathit{SPRPOS}(d,e)$ & Spread (first position $-$ last position) of mentions of $e$ in $d$ & \cite{Balog2013} & basic & N \\
		$\mathit{SIM}_{\mathit{cos}}(d, p_e)$ & Text cosine similarity between $d$ and $p_e$ & \cite{Balog2013} & basic & N \\
		$\mathit{SIM}_{\mathit{jac}}(d, p_e)$ & Text jaccard similarity between $d$ and $p_e$ & \cite{Balog2013} & basic & N \\		
		$\mathit{PREMENTION}_h(d, e)$ & Mention count of entity in the previous $h$ hour before document creation time of $d$ & \cite{Wang2013} & basic & N \\
		\midrule
		$\mathit{DOCLEN}_{chunk}(d)$ & Length of document in number of chunks & this paper & basic & N \\
		$\mathit{DOCLEN}_{sent}(d)$ & Length of document in number of sentences & this paper & basic & N \\
		$\mathit{ENTITYTYPE}(e)$ & Type of $e$ (PER, ORG, or FAC) & this paper & basic & C \\
		$\mathit{PROFILETYPE}(e)$ & Profile type: \emph{wiki},\emph{web}, or \emph{null} & this paper & basic & C \\
		$\mathit{PROFILELEN}(e)$ & Length of entity profile $e$ & this paper & basic & N \\
		$\mathit{ASPECTSIM}_k(d)$ & Cosine similarity between $d$ and $aspect_k$ estimated from Wikipedia & this paper & informativeness & N \\
		$\mathit{RELOPEN}_k(d)$ & Number of normalized open relation phrases $k$ in $d$ & this paper & informativeness & N \\
		$\mathit{RELSCHEMA}_k(d)$ & Number of relation type $k$ in document $d$ & this paper & informativeness & N \\\
		$\mathit{NUMENTITIES}(d)$ & Number of unique entity mentions in the documents & this paper & entity saliency & N \\
		$\mathit{NUMMENTIONS}(d)$ & Number of entity mentions in the documents & this paper & entity saliency & N \\
		$\mathit{NUMSENT}(d,e)$ & Number of sentences in $d$ containing entity $e$ & this paper & entity saliency & N \\
		$\mathit{FULLFRAC}(d,e)$ & Number of full mentions of $e$ in the document, normalized by number of entity mentions & this paper & entity saliency & N \\
		$\mathit{MENTIONFRAC}(d,e)$ & Number of full or partial mentions of $e$ in the document, normalized by number of entity mentions & this paper & entity saliency & N \\ 
		$\mathit{TMATCH}_Y(d)$ & Number of year expressions of timestamp $t$ in $d$ & this paper & timeliness & N \\
		$\mathit{TMATCH}_{\mathit{YM}}(d)$ & Number of year, month expressions of timestamp $t$ in $d$ & this paper & timeliness & N \\
		$\mathit{TMATCH}_{\mathit{YMD}}(d)$ & Number of year, month, date expressions of timestamp $t$ in $d$ & this paper & timeliness & N \\
		\bottomrule
	\end{tabularx}
\end{table*}

\paragraph{Closed relation extraction} The last informativeness feature is based on the occurrence of a set of pre-defined relations within the text of the candidate document. We obtain all relation mentions detected in the text by a relation extraction system, the Serif tagger~\cite{Boschee2005}. In our task, the corpus contains annotations of relation types based on the ACE relation schema~\cite{Doddington2004}. We only consider relations involving entities that are a person, organization, or location which amounts to 15 ACE relation types.
%
We construct a vector of the ACE relation types at the document level:
\begin{equation}
	S_k(d) = count(d, s_k),
\end{equation}
where $count(d,s)$ is the count of detected relations $k$ in the document. We refer to $S_k$ as the $\mathit{RELSCHEMA}$ features in Table~\ref{table:features}.

\subsubsection{Entity saliency features}

The entity saliency features $F_E$ aim to capture how prominently an entity features within a document. Although the basic features (defined in \S\ref{section:basicfeatures}) might capture some notion of saliency, they are focused on the target entity only. We extend this by looking at mentions of other entities within the document. For example, if $e$ is the only entity mentioned in the document then it is probably the main focus of the document. 

We define a \emph{full mention} as the complete name used to refer an entity in the document and a \emph{partial mention} as the first or last name of the entity. We introduce the following novel features based on this notion of entity saliency. The first feature is simply the number of entities in the document:
\begin{equation}
	\mathit{DOCENTITIES}(d) = \left|M\right|,
\end{equation}
where $M$ is the set of all entity mentions. 
The next feature is the number of entity mentions:
\begin{equation}
	\mathit{DOCMENTIONS}(d) = \sum_{e'}n(d,m_{e'}),
\end{equation}
that is, the total number of entity mentions as identified by the Serif tagger.
The next feature is the number of sentences containing the target entity $e$:
\begin{equation}
	\mathit{NUMSENT}(d, e) = \left|S_e\right|,
\end{equation}
where $S_e$ is the set of all sentences mentioning entity $e$.

We further define the fraction of full mentions of $e$ with respect to all entity mentions in the document:
\begin{equation}
	\mathit{FULLFRAC}(d,e) = \frac{n_{full}(d,m_e)}{\sum_{e'}n(d,m_{e'})},
\end{equation}
and also include the fraction of partial mentions $m_e$ of $e$ with respect to all entity mentions in the document:
\begin{equation}
	\mathit{MENTIONFRAC}(d,e) = \frac{n_{partial}(d,m_{e})}{\sum_{e'}n(d,m_{e'})},
\end{equation}
where $n(d,m)$ counts the number of mentions in document $d$ again obtained by the named entity recognizer.

\subsubsection{Timeliness features}

Timeliness features $F_T$ capture how timely a piece of information mentioned in the document is. 
We extract these features by comparing the document metadata containing the document creation time $t$ with the time expressions mentioned in the documents:
\begin{equation}
	\mathit{TMATCH}_Y(d) = count(year(t), d),
\end{equation}
where $count(year(t),d)$ counts the occurrences of year expressions of $t$ appear in the document. 
\begin{equation}
	\mathit{TMATCH}_{\mathit{YM}}(d) = count(yearmonth(t), d),
\end{equation}
where $\mathit{count}(\mathit{yearmonth}(t),d)$ counts the number of times year and month expressions of $t$ appear in the document. Finally,
\begin{equation}
	\mathit{TMATCH}_{\mathit{YMD}}(d) = count(yearmonthday(t), d),
\end{equation}
where $count(yearmonthday(t),d)$ counts the number of times the year, month, and date expressions of $t$ occur in the document $d$.

\subsection{Basic features}
\label{section:basicfeatures}

This section describes basic features $F_B$ that are commonly implemented in an entity-oriented document filtering system~\cite{Balog2013,Wang2013}, as described in \S\ref{section:related-work}. We also propose some new basic features.

\paragraph{Document features} Features extracted from document $d$, capturing the characteristics of $d$ independent of an entity. This includes the length, type, and language of $d$.

\paragraph{Entity features} Features based on knowledge about entity $e$ including, for instance, the number of related entities in the entity's profile $p_e$. In addition, we incorporate the length of profile $p_e$ and the type of entity profile available: \emph{Wiki}, \emph{Web}, or \emph{Null}.

\paragraph{Document-entity features} Features extracted from an entity and document pair. This includes the occurrences of full and partial mentions of $e$ in the document as well as the first and last position of occurring. They also include similarity between $d$ and $p_e$ and the number of related entities of $e$ mentioned in the document. 

\paragraph{Temporal features} Temporal features extracted from the occurrences of $e$ within the stream corpus $S$. After aggregating entity mentions in hourly bins, we obtain the counts in the previous $k$ hours before the creation of document $d$, where $k \in \{1,\ldots, 10\}$.

\subsection{Machine learning model} 

Next, we detail our classification-based machine learning model. We formulate the task as binary classification and train a classifier to distinguish vital and non-vital documents using the concatenated vector of all features described previously: $F = F_{B}\cup F_{I}\cup F_{E}\cup F_{T}$.
We train a global model $M$ in an \emph{entity-independent} way, utilizing all training data available for the model. Creating such a general model has the benefit that it can be readily applied to entities that do not exist in the training data.

We use gradient boosted decision trees (GBDT) \cite{Friedman2000} as our machine learning algorithm. GBDT learns an ensemble of trees with limited complexity in an additive fashion by iteratively learning models that aim to correct the residual error of previous iterations.
%
%
%
To obtain the probabilistic output as required by Eq.~\ref{eq:problem}, the gradient boosting classifier is trained as a series of weak learners in the form of regression trees. Each regression tree $t \in M$ is trained to minimize mean squared error on the logistic loss:
\begin{equation}
	\mathit{MSE} = \frac{1}{n} \sum_i^n \left(y_i^2 - \left(\frac{1}{1 + e^{pred_i}}\right)^2\right),
\end{equation}
where $y$ is the training label converted to either 0 or 1 for the negative and positive class, respectively, and $pred$ is the prediction score of the regression tree at data point $i$. The trees are trained in a residual fashion until convergence. 
At prediction time, each tree produces a score $s_t$; these are combined into a final score $s$, which is then converted into a probability using the logistic function:
\begin{equation}
	P = \frac{1}{1 + e^{-s}}.
\end{equation}
We take this output as our estimate of Eq.~\ref{eq:problem}. We refer to our proposed entity-independent document filtering method as {\OurApproach{}}.


\section{Experimental Setup}
\label{section:experimental-setup}

In this section we detail our experimental setup including the data that we use, the relevance assessments, and the evaluation metrics. Our experiments address the following research questions: 
  \begin{inparaenum}[(RQ1)]
  	\item How does our approach, \OurApproach{}, perform for vital document filtering of long-tail entities? 
	\label{RQ:1}
  	\item How does \OurApproach{} perform when filtering documents for entities not seen in the training data?
	\label{RQ:2}
	\item How does \OurApproach{} compare to the state-of-the-art for vital document filtering in terms of overall results?
	\label{RQ:3}
  \end{inparaenum}

\subsection{Data and annotations}

The TREC KBA StreamCorpus contains 1.2B documents. Rough\-ly half of these (579M) have been annotated with rich NLP annotations using the Serif tagger~\cite{Frank2014}. This annotated set is the official document set for TREC KBA 2014. Out of these annotated documents, a further selection is made for the Cumulative Citation Recommendation (CCR) task of KBA 2014. This results in  the final \emph{kba-2014-en-filtered} subset of 20,494,260 documents, which was filtered using surface form names and slot filling strings for the official query entities for KBA 2014. These documents are heterogeneous and originate from several Web sources: arxiv, classifieds, forums, mainstream news, memetracker, news, reviews, social, and blogs. We perform our experiments on this filtered subset.

The entities used as test topics are selected from a set of people, organizations, and facilities in specific geographical regions (Seattle, Washington, and Vancouver). The test entities consist of 86 people, 16 organizations, and 7 facilities, 74 of which are used for the vital document filtering task. Assessors judged $\sim$30K documents, which included most documents that mention a name from the handcrafted list of surface names of the 109 topic entities. Entities can have an initial profile in the form of \emph{wikipedia}, \emph{web}, or \emph{null}, indicating that no entity profile is given as a description of the entity. In order to have enough training data for each entity, the collection was split based on per-entity cut-off points in time. Some of the provided profile pages are dated after the training time cutoff of an entity. To avoid having access to future information, we filter out entity profiles belonging to those cases. Table~\ref{table:data2}  provides a breakdown of profile types of the test entities.

\begin{table}[t]
\caption{{Distribution of entity profile types and examples.}}
\label{table:data2}
\begin{tabularx}{\linewidth}{Xccl}
  \toprule
  \textbf{Entity profile} & \textbf{Count} & \textbf{Examples} \\
  \midrule
   \emph{Wiki} & 14 & \emph{Jeff Mangum, Paul Brandt} \\
   \emph{Web} & 19 & \emph{Anne Blair, Bill Templeton} \\
   \emph{Null} & 41 & \emph{Ted Sturdevant, Mark Lindquist} \\
  \bottomrule
 \end{tabularx}
\end{table}%

Annotators assessed entity-document pairs using four class labels: \emph{vital}, \emph{useful}, \emph{neutral}, and \emph{garbage}. For a document to be annotated as \emph{vital} means that the document contains \begin{inparaenum}[(1)] \item information that at the time it entered the stream would motivate an update to the entity's collection of key documents with a new slot value, or \item timely, new information about the entity's current state, actions, or situation.\end{inparaenum}{} Documents annotated as \emph{useful} are possibly citable but do not contain timely information about the entity. \emph{Neutral} documents are documents that are informative, but not citable, e.g., tertiary sources of information like Wikipedia pages. \emph{Garbage} documents are documents that are either spam or contain no mention of the entity. The distribution of the labels is detailed in Table~\ref{table:data1}. As our model performs binary classification, we collapse the non-vital labels into one class during training.

One of our proposed features is based on generic Wikipedia sections of \emph{Person} and \emph{Location} entities. For this purpose, we use a Wikipedia dump from January 2012.

\subsection{Experiments}
\label{section:experiments}
We run three experiments: two main experiments aimed at assessing the performance of \OurApproach{} on long-tail entities and on unseen entities, and a side experiment in which we determine the performance on all entities.

\paragraph{Main experiment: Long-tail entities} Our main experiment aims to answer RQ\ref{RQ:1} and adapts the standard TREC KBA setting with one difference: we aggregate the results for different entity popularity segments. {We define \emph{long-tail entities} to be entities without a Wikipedia or Web profile in the TREC KBA ground truth data.} All training entities are used to train the model and, during evaluation, a confidence score is assigned to every candidate document. All experiments are performed on the already pre-filtered documents using the canonical name of the entities as detailed above. Only documents containing at least a full match of the entity name are therefore considered as input. We focus on distinguishing vital and good documents, and use only documents belonging to these labels as our training data. 

\paragraph{Main experiment: Unseen entities} 
Our second main experiment aims to assess the performance of \OurApproach{} on \emph{unseen entities}, i.e., entities not found in the training data (RQ\ref{RQ:2}). We design this experiment as follows. We randomly split the query entities into five parts and divide the training data accordingly. For every iteration we train on the training data consisting only of document-entity pairs of the corresponding entity split and test on the remaining split. We perform this procedure five times, resulting in a 5-fold cross-validation. 

 \paragraph{Side experiment: All entities} Our side experiment aims to answer RQ\ref{RQ:3} and follows the standard TREC KBA setting. All entities within the test set are considered in the evaluation (i.e., the results are not segmented) to asses the overall performance of \OurApproach{}.

\begin{table}[t]
\caption{Label distribution in the ground truth.}
\label{table:data1}
\begin{tabularx}{\linewidth}{Xrr}
  \toprule
  \textbf{Label} & \textbf{Training} & \textbf{Test} \\
  \midrule
   \emph{Vital} & \num{1360} & \num{4665} \\
   \emph{Useful} &  \num{5482} & \num{20370} \\
   \emph{Neutral} &  \num{522} & \num{2044} \\
   \emph{Garbage} & \num{3302} & \num{1961} \\
  \bottomrule
 \end{tabularx}
\end{table}%

\subsection{Evaluation}

In our experiments, we use the evaluation metrics introduced in the TREC KBA track for the vital filtering task: $F_{macro}$, and maximum scaled utility ($SU$). We also compute precision ($P$), recall ($R$), and $F$ measure: the average of the harmonic mean of precision and recall over topics. For significance testing of the results, we use the paired t-test.

The main evaluation metric, $F_{macro}$, is defined as the \emph{maximum} of the harmonic mean of averaged precision and recall computed at every possible threshold $\theta$ which separates vital and non-vital documents: $\max(\avg(P),\avg(R))$. The motivation behind this is evaluation setup is as follows. A filtering system will have a single confidence threshold $\theta$ for which the classification performance is maximized. Different systems might have different optimal confidence score calibrations, hence choosing the maximum scores with respect to each system's best threshold would ensure the fairest comparison. 
Below we explicitly distinguish between $F_{macro}$ and $F$ when reporting our experimental results. 

$SU$ is a linear utility measure that assigns credit to the retrieved relevant and non-relevant documents and is computed as follows:
\begin{equation*}
  SU = \frac{\max(NormU, MinU) - MinU}{1 - MinU},
\end{equation*}
where $MinU$ is a tunable minimum utility (set to $-0.5$ by default), and $NormU$ is the normalized version of utility function $U$ which assigns two points for every relevant document retrieved and minus one point for every non-relevant document. The normalization is performed by dividing $NormU$ with the maximum utility score (i.e., 2 times the number of relevant documents). The official TREC KBA scorer sweeps over all the possible cutoff points and the reports the maximum $SU$. To gain additional insight, we also computed $SU$ at the cutoff $\theta$ with the best $F_{macro}$:
$SU_{\theta}$. 

\subsection{Baselines}

In our main experiments, we consider the following baseline approaches to compare the effectiveness of our approach.

\paragraph{Official Baseline \cite{Frank2014}} The official baseline in TREC KBA considers matched name fractions as the confidence score.

\paragraph{BIT-MSRA \cite{Wang2013}} A random forest, \emph{entity-independent} classification approach utilizing document, entity, document-entity, and temporal features. This approach achieved the best official performance at the TREC KBA 2013 track.

\smallskip\noindent%
In our side experiment aimed at assessing the performance of \OurApproach{} on all entities we also consider a state-of-the-art entity-dependent approach. 

\paragraph{MSR-KMG \cite{Jiang2014}} A random forest, \emph{entity-dependent} classification approach based on document cluster, temporal, entity title and profession features, with globally aligned confidence score. This approach achieved the best official performance in TREC KBA 2014. We take the team's best automatic run for comparison.

\subsection{Parameters and settings}
Recall that a document filtering system should output an estimate of $P(rel\mid d_e,e)$ (Eq.~\ref{eq:problem}).  The official KBA setup expects a confidence score in the $[0,1000]$ range for each decision made regarding a document. To make the initial output of our model compatible with this setup, the probabilities are mapped to a confidence score that falls in this interval by adopting the mapping procedure introduced in \cite{Balog2013}---we multiply the probability by 1000 and take the integer value.

Our approach involves two sets of hyperparameters. The first set deals with the machine learning algorithm of our choice. GBDT depends on two key parameters: the number of trees, $k$, and the maximum depth of each tree, $d$. The other set of parameters concerns the informativeness features. That is, the number of aspects that we used for the aspects-features, $m$, and the number of open relation patterns to consider, $n$. 

We perform cross-validation on the training data to select the values of these parameters. For the GDBT parameter we consider $k=[100,250,500]$ and tree depth $d=[6,7]$. For the informativeness parameters, we consider $m=[30,40,50]$ for the number of aspects and $n=[150,200,250]$ for number of the open relation patterns. We select the combination of parameters which maximize the mean F score across the validation folds, and finally set $k=100$, $d=6$, $m=50$, and $n=200$.


\newcommand{\duwtje}{\phantom{$^\blacktriangle$}}

\section{Results and Discussion}
\label{section:results}

In this section, we present and analyze our experimental results. 

\subsection{Main experiment: Long-tail entities}
 
One of our goals in this work is to develop methods that are specifically geared towards filtering documents for long-tail entities. Therefore, we are particularly interested in comparing the performance of the methods on entities with different levels of popularity.
To gain insight into our results along this dimension we segment the results by entity popularity using the type of entity profile as a proxy for popularity as defined in \S\ref{section:experiments}. We compute the best threshold for each approach, determine its per-entity performance using this cutoff, and then aggregate the performance by averaging the per-entity scores. We present these results in Table~\ref{table:experiment1}. Here, we answer RQ1 and compare our approach with other \emph{entity-independent} approaches. 

\begin{table}[t]
\caption{Results segmented by entity popularity. Significance of \OurApproach{} result is tested against the strong baseline (BIT-MSRA). Significant improvement is denoted with $^\blacktriangle$ ($p < 0.05$). Here the \emph{null profiles} segment represents the long-tail entities.}
\label{table:experiment1}
\begin{tabularx}{\linewidth}{Xcccc}
  \toprule
  \textbf{Segment} & \textbf{P} & \textbf{R} & \textbf{F} & \textbf{$SU_\theta$}\\
  \midrule
  \multicolumn{3}{l}{\emph{Null profiles}} \\
  Official baseline & 0.279\duwtje & \textbf{0.973}\duwtje & 0.388\duwtje & 0.268\duwtje \\
  BIT-MSRA & 0.362\duwtje & 0.630\duwtje & 0.404\duwtje & 0.313\duwtje \\
  \OurApproach{} & \textbf{0.398}$^\blacktriangle$ & 0.645\duwtje & \textbf{0.433}$^\blacktriangle$ & \textbf{0.350}$^\blacktriangle$ \\
  \midrule
  \multicolumn{3}{l}{\emph{Web profiles}} \\ 
  Official baseline & 0.391\duwtje & \textbf{1.000}\duwtje & 0.513\duwtje & 0.381\duwtje \\ 
  BIT-MSRA & \textbf{0.430}\duwtje & 0.867\duwtje & \textbf{0.536}\duwtje & \textbf{0.429}\duwtje \\
  \OurApproach{} & 0.424\duwtje & 0.827\duwtje & 0.517\duwtje & 0.410\duwtje \\
  \midrule
  \multicolumn{3}{l}{\emph{Wiki profiles}} \\ 
  Official baseline & 0.169\duwtje & \textbf{0.975} & 0.275\duwtje & 0.044\duwtje \\
  BIT-MSRA & 0.204\duwtje & 0.737 & 0.296\duwtje & 0.121\duwtje \\
  \OurApproach{} & \textbf{0.227}$^\blacktriangle$ & 0.704 & \textbf{0.317}\duwtje & \textbf{0.130}\duwtje \\
  \bottomrule
 \end{tabularx}
\end{table}%

First, we look at the average scores in each popularity group, starting with the \emph{Null} segment, which represents the long-tail entities in our setting. In the \emph{Null} segment, the recall performance of different methods is considerably lower than on the other two segments, but this is complemented by the fact that precision is higher than for the \emph{Wiki} segment. One important factor in this analysis is that these are most likely tail entities with very few candidate documents to consider.
More importantly, our approach achieves a significant improvement in the \emph{Null} segment, while keeping a comparable or better performance as compared to BIT-MSRA on the \emph{Wiki}, and \emph{Web} segments. In particular, the improvements in precision, $F$, and $SU_\theta$ in this segment are statistically significant.

\begin{figure}[h]
 \centering
    \subfigure[EIDF]
    {\includegraphics[clip=true,trim=0mm 0mm 0mm 15mm,width=\columnwidth]{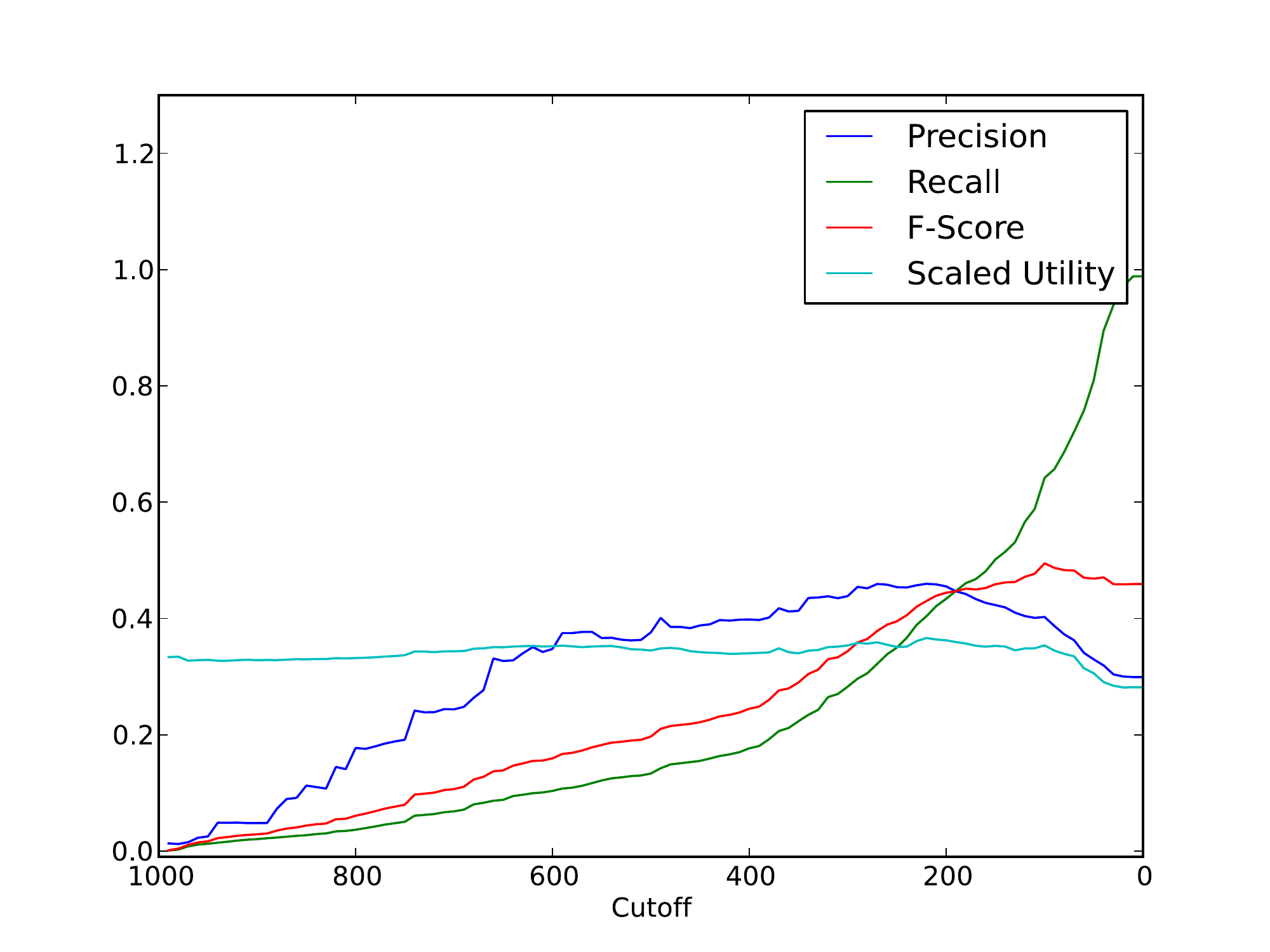}}
    \subfigure[BIT-MSRA]
    {\includegraphics[clip=true,trim=0mm 0mm 0mm 15mm,width=\columnwidth]{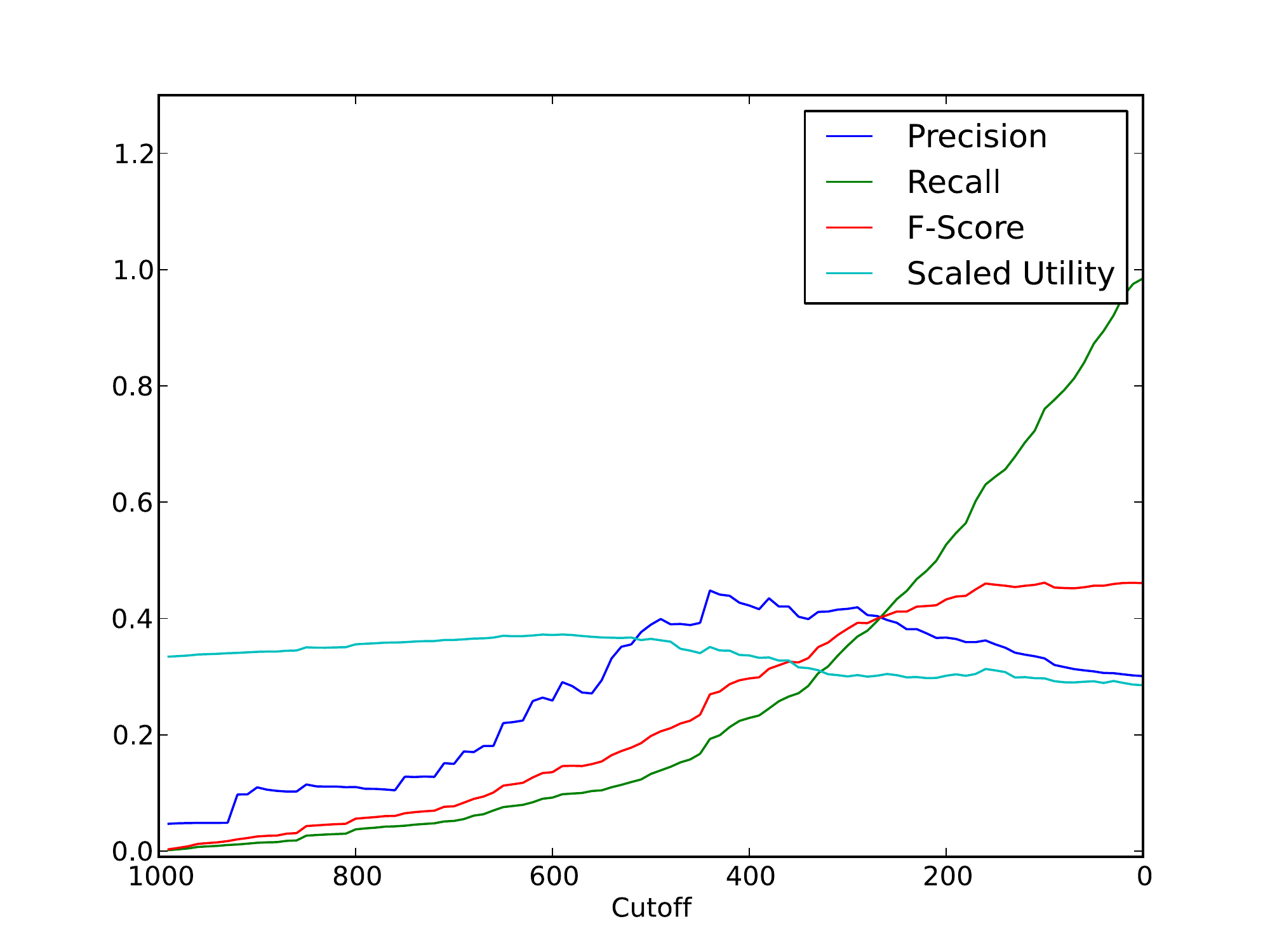}}
    \caption{Performance of EIDF and BIT-MSRA for long-tail entities across different cutoff points.}
  \label{fig:plot1}
  \label{fig:plot2}
\end{figure}

This finding is important because it confirms the effectiveness of our approach in the setting of long-tail entities. Faced with a considerably smaller pool of candidate documents in this segment, \OurApproach{} manages to detect more vital documents while simultaneously improving precision. Note that in the TREC KBA 2014 track, long-tail entities constitute a large fraction of the query entities (41 entities, i.e., 56\%). The performance of \OurApproach{} and BIT-MSRA for long-tail entities across different cutoff points is shown in Fig.~\ref{fig:plot1}.

Filtering documents for the \emph{Web} profile segment seems to be the easiest relative to the other segments. Recall and precision are highest compared to the other groups, which explains the higher $F$ score. Our approach, \OurApproach{}, achieves a $P$ score of 0.424, an $F$ score of 0.517 and $SU_\theta$ of 0.410 in this segment. This happens to be lower than the strong baseline (BIT-MSRA), but the differences in performance in this segment are not statistically significant.

Interestingly, the performance of all methods when filtering documents of entities belonging to the \emph{Wiki} group is the lowest. The recall is relatively high, but the $F$ score is brought down by the lower precision. This may be due to the fact that these popular entities have a much larger pool of candidate documents, making the filtering task difficult because a system has to recover only a selective fraction of the documents. Thus, faced with a large set of candidate documents, methods tend to work towards obtaining high recall. Despite this, \OurApproach{} manages to get the best precision, obtaining a significant improvement over the strong baseline. The low $SU_\theta$ scores indicate that it is difficult to beat a system that returns no relevant documents for this segment group. 

After looking at the general performance across the different segments, we compare the performance of our approach against the official TREC KBA baseline. Considerable gains are obtained in all three segments in terms of precision, $F$ and $SU_\theta$.

Informed by the previous insights, we also perform a follow-up experiment on training segment-conditioned models. Since feature value distributions might be different due to the popularity of an entity, we need to distinguish long-tail entities from more popular ones. One natural way of doing is to consider the existence of a knowledge base profile from Wikipedia---some entities may have a Wikipedia profile, some only an initial profile on a webpage, and some entities have no profile at all.
To capture this difference in characteristics, we train three separate machine learning models: $M_{wiki}$ for entities with a Wikipedia page, $M_{web}$ for entities with a lesser profile in the form of a Web page, and $M_{null}$ for entities with no profiles at all. During prediction, the appropriate model is automatically selected and applied to perform the predictions.
We failed to obtain any improvements with these segment-conditioned models. This may be due to the fact that by segmenting the data, we lose important information required to train our model with rich feature sets. To fully utilize the data while recognizing the different characteristics of each segment, a learning algorithm that can handle feature interaction, as we employ with tree-based ensembles, seems like a good solution. Having one global model that can handle feature interaction seems to be a better way to handle this problem, without resorting to individual models.

In sum, our approaches achieve the best performance overall across different segments, with the biggest performance gain realized for the long-tail entities segment. Importantly, the features designed for improvement in the long-tail entities segment do not have a significant detrimental effect on the results of other segments. In addition, learning a separate model for each segment does not yield additional benefits.

\subsection{Main experiment: Unseen entities}

In this section, we describe the results of our experiments on answering (RQ\ref{RQ:2}). The results of our experiments with unseen entities are detailed in Table~\ref{table:experiment6}. Our approach performs best on almost all folds in terms of $F_{macro}$, gaining significant improvements compared to other approaches on Fold1 and Fold3.

\begin{table}[t]
\caption{Results of cross-validation experiments with unseen entities, in terms of $F_{macro}$ (top), $P$ (middle), and $R$ (bottom).}
\label{table:experiment6}
\begin{tabularx}{\linewidth}{@{}l@{~~}X@{~~}X@{~~}X@{~~}X@{~~}X@{~~}l@{}}
  \toprule
  & \textbf{Fold1} & \textbf{Fold2} & \textbf{Fold3} & \textbf{Fold4} & \textbf{Fold5} & \mbox{}\hspace*{-1mm}\textbf{Overall}\\
  \midrule
  Official baseline & 0.410 & 0.482 & 0.401 & 0.532 & 0.400 & 0.445 \\
  BIT-MSRA & 0.405 & \textbf{0.489} & 0.413 & 0.537 & 0.407 & 0.450 \\
  \OurApproach{} & \textbf{0.458} & 0.485 & \textbf{0.438} & \textbf{0.539} & \textbf{0.408} & \textbf{0.465} \\
  \midrule
  Official baseline & 0.256 & 0.318 & 0.252 & 0.363 & 0.250 & 0.288 \\
  BIT-MSRA & 0.258 & \textbf{0.324} & 0.266 & 0.371 & 0.257 & 0.295 \\
  \OurApproach{} & \textbf{0.328} & 0.320 & \textbf{0.329} & \textbf{0.373} & 0.257 & \textbf{0.321} \\
  \midrule
  Official baseline & \textbf{1.000} & \textbf{1.000} & \textbf{0.975} & \textbf{0.993} & \textbf{1.000} & \textbf{0.994} \\
  BIT-MSRA & 0.956 & 0.992 & 0.923 & 0.972 & 0.976 & 0.964 \\
  \OurApproach{} & 0.762 & 0.996 & 0.654 & 0.973 & 0.987 & 0.874 \\
  \bottomrule
 \end{tabularx}
\end{table}%

Averaged over all folds, our approach also achieves the best performance. The differences between the performance of different methods in the unseen entities setting is very small in terms $F_{macro}$. Overall, the learned model tends to be precision-oriented with some loss in recall. Compared to the results of the main experiments (Table~\ref{table:experiment1}), the result is lower in terms of absolute score. 
This might be explained as follows. First, The model is now learning on less data---roughly 80\% of the full data, depending on the number of data points that contribute to the folds. Secondly, the model is now performing predictions on entities that may have very different characteristics than the ones found in the training data.
The average scores in each fold also vary considerably. This can be explained by the fact that by splitting the data in terms of entities, we might end up with different numbers of training and testing data in each split. Additionally, the inherent difficulty of filtering documents within each fold will also vary based on the popularity and the size of the candidate document pools.
The magnitude of the improvements obtained in each fold also tends to be smaller, because, with 80\% of the data, there are fewer positive examples available to learn a rich set of features (due to the imbalance of \emph{vital} and \emph{non-vital} document labels).

The results of filtering documents for unseen entities are quite promising, and the fact that the learning algorithm is able to achieve a better score than a name fraction baseline indicates that it is successful in learning the characteristics of vital documents and applying it to new, unseen entities.

\subsection{Side experiment: All entities}
To answer (RQ\ref{RQ:3}), we compare our method, \OurApproach, with \emph{entity-independent} and \emph{entity-dependent} baselines in terms of overall, non-segmented results. 
Table~\ref{table:experiment2} shows the results for this experiment. First, looking at the absolute scores, all methods improve over the official baseline in terms of $F_{macro}$, $SU$, and $P$. The official baseline unsurprisingly achieves the highest recall as it simply considers all document containing exact mentions of the target entity as vital.

\begin{table}[t]
\caption{Overall results with official and additional metrics. Significance of \OurApproach{} result is tested against the strong baseline (BIT-MSRA). Significant improvements are denoted with $^\blacktriangle$ ($p < 0.05$). The official TREC KBA scorer returns $F_{macro}$, $SU$, $P$, and $R$. We also compute additional metrics, $F$ and $SU_\theta$ to gain more insight about the results. We can not compute the significance test against MSR-KMG because the run is not available. Due to the way $F_{macro}$ is computed in TREC KBA, as a harmonic mean over recall and precision macro statistics, significance testing cannot be applied to $F_{macro}$.}
\label{table:experiment2}
\begin{tabularx}{\linewidth}{@{}X@{~~}c@{~~}c@{~~}c@{~~}c@{~~}c@{~~}c@{}}
  \toprule
  \textbf{Method} & \textbf{P} & \textbf{R} & \textbf{F} & \textbf{$SU_{\theta}$} & \textbf{$F_{macro}$} & \textbf{SU}\\
   \midrule
    \multicolumn{7}{@{}l}{\emph{Entity-independent}} \\ 
   Official baseline & 0.286\duwtje & \bf 0.980\duwtje & 0.397\duwtje & 0.253\duwtje & 0.442 & 0.333 \\
   BIT-MSRA & 0.348\duwtje & 0.709\duwtje & 0.415\duwtje & 0.305\duwtje & 0.467 & 0.370 \\
   \OurApproach{}  & 0.371$^\blacktriangle$ & 0.701\duwtje & 0.432$^\blacktriangle$ & 0.323$^\blacktriangle$ & 0.486 & 0.367 \\
   \midrule
    \multicolumn{7}{@{}l}{\emph{Entity-dependent}} \\ 
   MSR-KMG (automatic) \cite{Jiang2014} & \bf 0.378\duwtje & 0.744\duwtje & -- & -- & \bf 0.501 & \bf 0.377 \\
  \bottomrule
 \end{tabularx}
\end{table}%

Our approach also outperforms the two entity-independent baselines in terms of $F_{macro}$; we achieve significant improvements over BIT-MSRA in terms of precision, while maintaining the same level of recall. BIT-MSRA achieves a slightly better performance than EIDF in terms of $SU$. However, the difference is very small and not significant.

Compared to the best entity-dependent approach, \OurApproach{} obtains a comparable level of precision and $F_{macro}$. In summary, \OurApproach{} achieves the best entity-independent performance and competitive performance to the state of the art entity-dependent approach.

\subsection{Feature analysis}

Recall that we learn a single, entity-independent model across all entities.
We zoom in on the effectiveness of each feature within this global, entity-independent model. The importance of each feature is determined by averaging its importance across the trees that comprise the ensemble model.
\begin{table}[t]
\caption{Feature importance analysis for the model learned in the main and side experiments on long-tail entities.}
\label{table:experiment3}
\begin{tabularx}{\linewidth}{Xc}
  \toprule
  \textbf{Feature} & \textbf{Importance} \\
  \midrule
   $\mathit{FPOSFULL}(d,e)$ & 0.030 \\
   $\mathit{PROFILELEN}(e)$ & 0.025 \\
   $\mathit{FPOSFULL}_N(d,e)$ & 0.022 \\
   $\mathit{REL}(e)$ & 0.021 \\
   $\mathit{ASPECTSIM}_{\mathit{filmography}}(d)$ & 0.019 \\
   $\mathit{DOCLEN}_{\mathit{SENT}}(d)$ & 0.018 \\
   $\mathit{MENTIONFRAC}(d,e)$ & 0.016 \\
   $\mathit{PREMENTION}_{h2}(d,e)$ & 0.016 \\
   $\mathit{SIM}_{\mathit{cos}}(d,p_e)$ & 0.015 \\
   $\mathit{ASPECTSIM}_{\mathit{coachingcareer}}(d)$ & 0.015 \\
   $\mathit{LPOSFULL}(d,e)$ & 0.014 \\
   $\mathit{ASPECTSIM}_{\mathit{politicalcareer}}(d)$ & 0.013 \\
   $\mathit{LSPRFULL}_N(d,e)$ & 0.013 \\
   $\mathit{TMATCH}_Y(d)$ & 0.012 \\
   $\mathit{LPOSFULL}_N(d,e)$ & 0.012 \\
   $\mathit{SIM}_{\mathit{jaccard}}(d,p_e)$ & 0.012 \\
  \bottomrule
 \end{tabularx}
\end{table}%
We observe several things. First, the most important features are a combination of common features in document filtering, e.g., the first position of the entity, the spread of entity mentions, and our proposed features. One of our proposed features (profile length) is the most discriminative feature and another of our proposed saliency features, the fraction of entity mentions, is also shown to be quite important.  
As for the rest, the aspect-based features seem to be the most important informativeness features, with as many as three features belonging to the aspect-based group in the top most important features.

The aspect-based features might be complementary to the more common cosine and jaccard profile similarity features. In combination with the profile length feature the aspect-based features seem to be triggered when the profile similarity scores are zero, which will happen in the case of entities without a profile.
Having established this, we zoom in on the most important aspect-based features as detailed in Table~\ref{table:experiment4}. Recall that in our experiments, we use the top-50 aspects constructed from Wikipedia. Often, including aspects-based features seems intuitive, as is the case for, e.g., \emph{achievements}, \emph{accomplishment}, \emph{coaching-career}, and \emph{political-ca\-reer}, since they are things that are typically included in vital documents. 

\begin{table}[t]
\caption{Top Wikipedia aspect importance.}
\label{table:experiment4}
\begin{tabularx}{\linewidth}{Xc}
  \toprule
  \textbf{Feature} & \textbf{Importance} \\
  \midrule
   \emph{filmography} & 0.019 \\
   \emph{coaching-career} & 0.015 \\ 
   \emph{political-career} & 0.013 \\
   \emph{wrestling} & 0.011 \\
   \emph{references} & 0.011 \\
   \emph{championships-accomplishments} & 0.011 \\
   \emph{footnotes} & 0.011 \\
   \emph{achievements} & 0.011 \\
   \emph{selected-publications} & 0.010 \\
   \emph{links} & 0.010 \\
  \bottomrule
 \end{tabularx}
\end{table}%

All in all, we extracted 358 features. A breakdown of feature types in the top-30 features is shown in Table~\ref{table:experiment4b}. The informativeness features not ranked among the top in the table are not as discriminative as the Wikipedia aspects. In the case of open relation patterns, some receive a zero relative importance score. One possible explanation is that these patterns are very common and may occur in many documents, thus having very little discriminative power. In other cases, the patterns are quite rare, and they might thus only occur in a few documents. 

\begin{table}[t]
\caption{Feature types within the top-30.}
\label{table:experiment4b}
\begin{tabularx}{\linewidth}{Xc}
  \toprule
  \textbf{Feature type} & \textbf{Number of features} \\
  \midrule
   \emph{basic} & 14 \\
   \emph{informativeness} & 13 \\ 
   \emph{entity saliency} & 2 \\
   \emph{timeliness} & 1 \\
 \bottomrule
 \end{tabularx}
\end{table}%

\section{Conclusion and Future Work}
\label{section:conclusion}

In this paper we have addressed an information filtering task for long-tail entities on a stream of documents. In particular, we have developed and evaluated a method called EIDF for classifying vital and non-vital documents with respect to a given entity. We have done so by designing intrinsic features that capture the notions of \emph{informativeness}, \emph{entity saliency}, and \emph{timeliness} of documents. We have also considered the challenges related to filtering long-tail entities and have adjusted our features accordingly.
%
%
We have applied these features in combination with a set of basic document filtering features from the literature to train an \emph{entity-indepen\-dent} model that is also able to perform filtering for entities not found in the training data. 
%
Upon segmenting our results by entity popularity, as approximated by its profile type, we have found that our approach is particularly good at improving document filtering performance for long-tail entities. 
When looking at the overall results of experiments conducted on the TREC KBA 2014 test collection we have found that our approach is able to achieve competitive performance compared to state-of-the-art automatic \emph{entity-dependent} approaches. 
On filtering documents for unseen entities, we have found that our approach achieves a lower absolute performance overall than on seen entities, as is to be expected, but still improves over a strong name matching and classification baseline.
A feature analysis revealed two things. First, entity popularity, proxied using the profile length feature is important.
Second, informativeness features, and in particular aspect-based features derived from Wikipedia, are important for this task.

In summary, our results confirm the effectiveness of our entity-indepen\-dent document filtering approach for knowledge base acceleration for long-tail entities, with (1)~its ability to improve filtering performance specifically on the segment of tail entities, and (2)~its relatively good performance on classifying documents for unseen entities, i.e., those not found in the training data.

As to future work, we are interested in exploring several directions. First, it would be interesting to explore the effect of combining the proposed features with other machine learning algorithms. Our preliminary experiment in this direction with applying logistic regression as the underlying learning algorithm indicates that we can obtain similar improvements. 
Next, we aim to apply more semantic approaches such as entity linking to detect entities and concepts mentioned in the context of a target entity.
Last, we want to apply incremental learning so as to obtain a document filtering model that is able to learn from its previous decisions.

\smallskip

\begin{spacing}{1}
\noindent\small
\textbf{Acknowledgments.}
This research was supported by
Ahold,
Amsterdam Data Science,
Blendle,
the Bloomberg Research Grant program,
the Dutch national program COMMIT,
Elsevier,
the European Community's Seventh Framework Programme (FP7/2007-2013) under
grant agreement nr 312827 (VOX-Pol),
the ESF Research Network Program ELIAS,
the Royal Dutch Academy of Sciences (KNAW) under the Elite Network Shifts project,
the Microsoft Research Ph.D.\ program,
the Netherlands eScience Center under project number 027.012.105,
the Netherlands Institute for Sound and Vision,
the Netherlands Organisation for Scientific Research (NWO)
under pro\-ject nrs
727.\-011.\-005, 
612.001.116, 
HOR-11-10, 
640.006.013, 
612.\-066.\-930, 
CI-14-25, 
SH-322-15, 
652.\-002.\-001, 
612.\-001.\-551, 
652.\-001.\-003, 
the Yahoo Faculty Research and Engagement Program,
and
Yandex.
All content represents the opinion of the authors, which is not necessarily shared or endorsed by their respective employers and/or sponsors.
\end{spacing}

\vspace*{-.5\baselineskip}
\providecommand{\bibfont}{\small}
\renewcommand{\bibsection}{\section*{REFERENCES}}
\setlength{\bibsep}{2pt}
\bibliographystyle{abbrvnatnourl}
{\raggedright\small
\bibliography{cikm2016-kba}

\begin{thebibliography}{34}
\providecommand{\natexlab}[1]{#1}
\providecommand{\url}[1]{\texttt{#1}}
\expandafter\ifx\csname urlstyle\endcsname\relax
  \providecommand{\doi}[1]{doi: #1}\else
  \providecommand{\doi}{doi: \begingroup \urlstyle{rm}\Url}\fi

\bibitem[Allan(2002)]{Allan2002}
J.~Allan.
\newblock Introduction to topic detection and tracking.
\newblock In \emph{Topic Detection and Tracking}, pages 1--16. Kluwer Academic
  Publishers, 2002.

\bibitem[Balasubramanian and Cucerzan(2009)]{Balasubramanian2009}
N.~Balasubramanian and S.~Cucerzan.
\newblock Automatic generation of topic pages using query-based aspect models.
\newblock In \emph{CIKM '09'}, pages 2049--2052. ACM, 2009.

\bibitem[Balasubramanian and Cucerzan(2010)]{Balasubramanian2010}
N.~Balasubramanian and S.~Cucerzan.
\newblock Topic pages: An alternative to the ten blue links.
\newblock In \emph{ICSC '10}. IEEE, 2010.

\bibitem[Balog and Ramampiaro(2013)]{Balog2012}
K.~Balog and H.~Ramampiaro.
\newblock Cumulative citation recommendation: Classification vs. ranking.
\newblock In \emph{SIGIR '13'}, pages 941--944. ACM, 2013.

\bibitem[Balog et~al.(2013)Balog, Ramampiaro, Takhirov, and
  N{\o}rv{\aa}g]{Balog2013}
K.~Balog, H.~Ramampiaro, N.~Takhirov, and K.~N{\o}rv{\aa}g.
\newblock Multi-step classification approaches to cumulative citation
  recommendation.
\newblock In \emph{OAIR '13}, pages 121--128. Le Centre De Hautes Etudes
  Internationales D'Informatique Documentaire, 2013.

\bibitem[Banerjee and Mitra(2015)]{Banerjee2015}
S.~Banerjee and P.~Mitra.
\newblock {WikiKreator: Improving Wikipedia Stubs Automatically}.
\newblock In \emph{ACL '15}, pages 867--877, 2015.

\bibitem[Bonnefoy et~al.(2013)Bonnefoy, Bouvier, and Bellot]{Bonnefoy2013}
L.~Bonnefoy, V.~Bouvier, and P.~Bellot.
\newblock A weakly-supervised detection of entity central documents in a
  stream.
\newblock In \emph{SIGIR '13}. ACM, 2013.

\bibitem[Boschee et~al.(2005)Boschee, Weischedel, and Zamanian]{Boschee2005}
E.~Boschee, R.~Weischedel, and A.~Zamanian.
\newblock {Automatic information extraction}.
\newblock \emph{ICIA '05}, pages 2--4, 2005.

\bibitem[Cano et~al.(2014)Cano, Cs, Edu, and Cs]{Cano2014}
I.~Cano, I.~Cs, W.~Edu, and G.~Cs.
\newblock Distributed non-parametric representations for vital filtering: {UW
  at TREC KBA 2014}.
\newblock In \emph{TREC 2014}. NIST, 2014.

\bibitem[Cheng et~al.(2015)Cheng, Xu, and Qu]{Cheng2015}
G.~Cheng, D.~Xu, and Y.~Qu.
\newblock Summarizing entity descriptions for effective and efficient
  human-centered entity linking.
\newblock In \emph{WWW '15}, pages 184--194. International World Wide Web
  Conferences Steering Committee, 2015.

\bibitem[Dietz and Dalton(2013)]{Dietz2013}
L.~Dietz and J.~Dalton.
\newblock {UMass at TREC 2013 Knowledge Base Acceleration track}.
\newblock In \emph{TREC 2013}. NIST, 2013.

\bibitem[Doddington et~al.(2004)Doddington, Mitchell, Przybocki, Ramshaw,
  Strassel, and Weischedel]{Doddington2004}
G.~R. Doddington, A.~Mitchell, M.~A. Przybocki, L.~A. Ramshaw, S.~Strassel, and
  R.~M. Weischedel.
\newblock The automatic content extraction ({ACE}) program-tasks, data, and
  evaluation.
\newblock In \emph{LREC}, 2004.

\bibitem[Dunietz and Gillick(2014)]{Dunietz2014}
J.~Dunietz and D.~Gillick.
\newblock A new entity salience task with millions of training examples.
\newblock In \emph{EACL '14}. ACL, 2014.

\bibitem[Etzioni et~al.(2011)Etzioni, Fader, Christensen, Soderland, and
  Mausam]{Etzioni2011}
O.~Etzioni, A.~Fader, J.~Christensen, S.~Soderland, and Mausam.
\newblock {Open information extraction: The second generation}.
\newblock \emph{IJCAI 2011}, pages 3--10, 2011.

\bibitem[Fader et~al.(2011)Fader, Soderland, and Etzioni]{Fader2011}
A.~Fader, S.~Soderland, and O.~Etzioni.
\newblock Identifying relations for open information extraction.
\newblock In \emph{EMNLP '11}, 2011.

\bibitem[Fetahu et~al.(2015)Fetahu, Markert, and Anand]{Fetahu2015}
B.~Fetahu, K.~Markert, and A.~Anand.
\newblock Automated news suggestions for populating wikipedia entity pages.
\newblock In \emph{CIKM '15}, pages 323--332. ACM, 2015.

\bibitem[Fissaha~Adafre and de~Rijke(2005)]{fissaha-adafre-discovering-2005}
S.~Fissaha~Adafre and M.~de~Rijke.
\newblock Discovering missing links in wikipedia.
\newblock In \emph{Proceedings of the Workshop on Link Discovery: Issues,
  Approaches and Applications (LinkKDD-2005)}. ACM, August 2005.

\bibitem[Frank et~al.(2012)Frank, Kleiman-Weiner, Roberts, Niu, Ce,
  Christopher, and Soboroff]{Frank2012}
J.~R. Frank, M.~Kleiman-Weiner, D.~A. Roberts, F.~Niu, Z.~Ce, R.~Christopher,
  and I.~Soboroff.
\newblock Building an entity-centric stream filtering test collection for {TREC
  2012}.
\newblock In \emph{TREC 2012}. NIST, 2012.

\bibitem[Frank et~al.(2014)Frank, Kleiman-Weiner, Roberts, Voorhees, and
  Soboroff]{Frank2014}
J.~R. Frank, M.~Kleiman-Weiner, D.~A. Roberts, E.~Voorhees, and I.~Soboroff.
\newblock {TREC KBA Overview}.
\newblock In \emph{TREC 2014}. NIST, 2014.

\bibitem[Friedman(2000)]{Friedman2000}
J.~H. Friedman.
\newblock Greedy function approximation: A gradient boosting machine.
\newblock \emph{Annals of Statistics}, 29:\penalty0 1189--1232, 2000.

\bibitem[Gebremeskel and de~Vries(2014)]{Gebremeskel2014}
G.~G. Gebremeskel and A.~P. de~Vries.
\newblock Entity-centric stream filtering and ranking: Filtering and
  unfilterable documents.
\newblock In \emph{TREC 2014}, 2014.

\bibitem[Jiang and Lin(2014)]{Jiang2014}
J.~Jiang and C.-Y. Lin.
\newblock {MSR KMG at TREC 2014 KBA Track Vital Filtering Task}.
\newblock In \emph{TREC}, 2014.

\bibitem[Li et~al.(2011)Li, Wang, Gao, and Jiang]{Li2011}
P.~Li, Y.~Wang, W.~Gao, and J.~Jiang.
\newblock Generating aspect-oriented multi-document summarization with
  event-aspect model.
\newblock In \emph{EMNLP '11}, pages 1137--1146. ACL, 2011.

\bibitem[Liu et~al.(2013)Liu, Darko, and Fang]{Liu2013}
X.~Liu, J.~Darko, and H.~Fang.
\newblock A related entity based approach for knowledge base acceleration.
\newblock In \emph{TREC 2013}. NIST, 2013.

\bibitem[Pantel and Fuxman(2011)]{Pantel2011}
P.~Pantel and A.~Fuxman.
\newblock Jigs and lures: Associating web queries with structured entities.
\newblock In \emph{HTL '11}, 2011.

\bibitem[Reinanda and de~Rijke(2014)]{reinanda-prior-informed-2014}
R.~Reinanda and M.~de~Rijke.
\newblock Prior-informed distant supervision for temporal evidence
  classification.
\newblock In \emph{Coling '14}, pages 996--1006. ACL, August 2014.

\bibitem[Reinanda et~al.(2015)Reinanda, Meij, and
  de~Rijke]{reinanda-mining-2015}
R.~Reinanda, E.~Meij, and M.~de~Rijke.
\newblock Mining, ranking and recommending entity aspects.
\newblock In \emph{SIGIR '15}, pages 263--272. ACM, August 2015.

\bibitem[Song et~al.(2012)Song, Yu, Xu, Liu, Li, and Wen]{Song2012}
W.~Song, Q.~Yu, Z.~Xu, T.~Liu, S.~Li, and J.-R. Wen.
\newblock Multi-aspect query summarization by composite query.
\newblock In \emph{SIGIR '12}, pages 325--334. ACM, 2012.

\bibitem[Taneva and Weikum(2013)]{Taneva2013}
B.~Taneva and G.~Weikum.
\newblock Gem-based entity-knowledge maintenance.
\newblock In \emph{CIKM '13}, pages 149--158. ACM, 2013.

\bibitem[Voskarides et~al.(2015)Voskarides, Meij, Tsagkias, de~Rijke, and
  Weerkamp]{Voskarides2015}
N.~Voskarides, E.~Meij, M.~Tsagkias, M.~de~Rijke, and W.~Weerkamp.
\newblock Learning to explain entity relationships in knowledge graphs.
\newblock In \emph{ACL-IJCNLP 2015}, 2015.

\bibitem[Wang et~al.(2013)Wang, Song, Lin, and Liao]{Wang2013}
J.~Wang, D.~Song, C.~Lin, and L.~Liao.
\newblock {BIT and MSRA at TREC KBA CCR Track 2013}.
\newblock In \emph{TREC 2013}. NIST, 2013.

\bibitem[Wang et~al.(2015{\natexlab{a}})Wang, Song, Wang, Zhang, Si, Liao, and
  Lin]{Wang2015}
J.~Wang, D.~Song, Q.~Wang, Z.~Zhang, L.~Si, L.~Liao, and C.-Y. Lin.
\newblock An entity class-dependent discriminative mixture model for cumulative
  citation recommendation.
\newblock In \emph{SIGIR '15}, pages 635--644. ACM, 2015{\natexlab{a}}.

\bibitem[Wang et~al.(2015{\natexlab{b}})Wang, Song, Zhang, Liao, Si, and
  Lin]{Wang2015b}
J.~Wang, D.~Song, Z.~Zhang, L.~Liao, L.~Si, and C.-y. Lin.
\newblock {LDTM}: A latent document type model for cumulative citation
  recommendation.
\newblock In \emph{EMNLP '15}, pages 561--566. ACL, 2015{\natexlab{b}}.

\bibitem[Zhou and Chang(2013)]{Zhou2013}
M.~Zhou and K.~C.-C. Chang.
\newblock Entity-centric document filtering: Boosting feature mapping through
  meta-features.
\newblock In \emph{CIKM '13}, pages 119--128. ACM, 2013.

\end{thebibliography}
}

\end{document}